\documentclass[prd,nofootinbib,preprint,superscriptaddress]{revtex4}

\usepackage{amsmath, amssymb, amsthm, graphicx, epsfig, fancyhdr,epsfig, slashed,bm}

\usepackage{tikzsymbols}
\usepackage{natbib}
\usepackage{float}

\usepackage{tikz,xcolor,hyperref}

\definecolor{lime}{HTML}{A6CE39}
\DeclareRobustCommand{\orcidicon}{
	\begin{tikzpicture}
	\draw[lime, fill=lime] (0,0) 
	circle [radius=0.2] 
	node[white] {{\fontfamily{qag}\selectfont \tiny ID}};
	\draw[white, fill=white] (-0.0625,0.095) 
	circle [radius=0.007];
	\end{tikzpicture}
	\hspace{-2mm}
}

\foreach \x in {A, ..., Z}{\expandafter\xdef\csname orcid\x\endcsname{\noexpand\href{https://orcid.org/\csname orcidauthor\x\endcsname}
			{\noexpand\orcidicon}}
}

\newcommand{\be}{\begin{equation}}
\newcommand{\ee}{\end{equation}}
\newcommand{\bea}{\begin{eqnarray}}
\newcommand{\eea}{\end{eqnarray}}

\begin{document}

%\title{Confinement Conditions and $\beta$-functions in String-inspired \\ Infinite Derivative Non-local  Gauge Theories}

\title{Non-perturbative Origin of the Electroweak Scale: \\ \it{RGE in Strongly-coupled Dark Gauge Theories via Dyson-Schwinger} }

\author{Marco Frasca\orcidA{}}
\email{marcofrasca@mclink.it}
\affiliation{Rome, Italy}

\author{Anish Ghoshal\orcidB{}}
\email{anish.ghoshal@fuw.edu.pl}
\affiliation{Institute of Theoretical Physics, Faculty of Physics, University of Warsaw, ul. Pasteura 5, 02-093 Warsaw, Poland}

\author{Nobuchika Okada\orcidC{}}
\email{okadan@ua.edu}
\affiliation{Department of Physics and Astronomy, \\ University of Alabama, Tuscaloosa, AL 35487, USA}

\begin{abstract}
\textit{We propose a novel pathway to generate the electroweak scale (EW) via non-perturbative dynamics of 
a dark gauge sector based on the SU(N) gauge group. 
Imposing the scale invariance of the theory, we investigate the electroweak symmetry breaking (EWSB) which is triggered dynamically via the condensation of gauge fields. 
% Added after refere's comment
Instead of the usual dimension-4 triggered breaking, a dimension 6 term (with Wilson coefficient compatible with SMEFT bounds) coupling the Higgs boson and the Yang-Mills field quadratic term in the Lagrangian provides feedback from the gauge to the Higgs sector.
We provide a novel method to estimate a non-perturbative EW scale generation using 
% Modified after referee's comments
%the exact solution of the background equations of motion 
the exact solution of a truncated set of the background equations of motion
in Yang-Mills theory in terms of Jacobi elliptic functions and the exact beta-function valid in the strongly coupled regimes via the Dyson-Schwinger approach. Particularly, we find an analytical result for the Renormalization Group Equation (RGE) of the gauge coupling in the $SU(N)$ sector in the strongly-coupled regime. The dynamics studied in this paper pave the way to a more realistic model building with possible resolution to the hierarchy problem and, in general, dynamical generation of scales. 
}
\end{abstract}

\maketitle

\section{Introduction}

With null results at the Large Hadron Collider (LHC) for any evidence of supersymmetry, alternatives to the solution of the gauge hierarchy problem are being explored. An intriguing possibility is that the hierarchy problem might be related to the scale invariance at the classical level, which is only broken at the quantum level due to quantum anomalies. As has been first shown, the dynamical generation of the gauge symmetry breaking is realized via radiative symmetry breaking by the Coleman-Weinberg mechanism \cite{Coleman:1973jx}.  However, when the mechanism is applied to the Standard Model (SM), the mass of the gauge bosons is predicted to be greater than that of the Higgs boson, $m_{Z,W} > m_H$. In the framework of Beyond the SM (BSM), realistic dynamical generations of the electroweak (EW) scale have been explored extensively in the literature \cite{Adler:1982ri,Salvio:2014soa,Einhorn:2014gfa,Einhorn:2016mws,Einhorn:2015lzy,Englert:2013gz,Holthausen:2009uc,Meissner:2006zh,Foot:2007as,Farzinnia:2013pga}. Moreover, in the context of non-minimally coupled gravity, scale invariant models naturally provide flat inflationary potentials \cite{Khoze:2013uia,Kannike:2014mia,Rinaldi:2014gha,Salvio:2014soa,Kannike:2015apa,Kannike:2015fom,Barrie:2016rnv,Tambalo:2016eqr}, and also the mass scale of dark matter (DM) can be dynamically generated \cite{Hambye:2013sna,Karam:2015jta,Kannike:2015apa,Kannike:2016bny,Karam:2016rsz}. This has always been seen as a direction of model-building toward the dynamically generated solution to the gauge hierarchy problem in the SM
\cite{Foot:2007iy,AlexanderNunneley:2010nw,Englert:2013gz,Hambye:2013sna,Farzinnia:2013pga,Altmannshofer:2014vra,Holthausen:2013ota,Salvio:2014soa,Einhorn:2014gfa,Kannike:2015apa,Farzinnia:2015fka,Kannike:2016bny}. See Refs. \cite{1306.2329,1410.1817,0902.4050,0909.0128,1210.2848,1703.10924,1807.11490} for other studies of conformal invariance and dimensional transmutation of energy scales \cite{2012.11608,1812.01441}. 

%Refs.~\cite{Jaeckel:2016jlh, Jinno:2016knw, Marzola:2017jzl, Iso:2017uuu, Chao:2017ilw, Baldes:2018emh, Prokopec:2018tnq, Brdar:2018num, Marzo:2018nov, Hasegawa:2019amx, Ghoshal:2020vud, Ellis:2020nnr,  Chikkaballi:2023cce, Ahriche:2023jdq} for exploitations of this mechanism in other BSM contexts. This has also been used in the context of baryogenesis via leptogenesis~\cite{Huang:2022vkf,Dasgupta:2022isg,Borah:2022cdx}, complementarity with collider searches~\cite{Dasgupta:2023zrh} and of the generation of the Planck scale~\cite{Ghoshal:2022qxk}.

Besides the Coleman-Weinberg pathway, dynamical generation of mass scales in scale-invariant theories can be achieved via dimensional transmutation in a strongly interacting sector based on hidden gauge groups \cite{Holthausen:2013ota,Hambye:2007vf,Hur:2011sv,Hambye:2013dgv}, including some examples from standard quantum chromodynamics (QCD) (see e.g.~\cite{Kubo:2014ova}). This is purely due to non-perturbative effects. Already in the 1980s, the authors of \cite{Marciano:1980zf,Zoupanos:1983xh,Lust:1985aw,Lust:1985jk} suggested to dynamically break the EW symmetry by the condensation of chiral fermions in high color representations, see Ref. \cite{Klett:2022iga} for recent consideration. 

In this paper, we propose a scale-invariant scenario with a dark sector based on the SU(N) gauge group, in which the electroweak symmetry breaking (EWSB) is triggered dynamically via the condensation of gauge fields. This mechanism enables us to understand the dynamic generation of the EW scale. The main part of this work is devoted to the development of 
% Removed after referee's comment
%a novel technique involving 
%
a novel method to estimate the non-perturbative EW scale generation based on the exact solution of the background equations of motion in Yang-Mills theory in terms of Jacobi elliptic functions and the exact beta-function valid in the strongly-coupled regimes via Dyson-Schwinger approach. Particularly, we find an analytical result for the Renormalization Group (RG) running of the gauge coupling in the classically conformal SU(N) sector in the strongly-coupled regime.
To study the non-perturbative regime, we utilize the exact solutions found in terms of the Jacobi elliptical functions following the analytic approach of Dyson-Schwinger equations, originally devised by Bender, Milton and Savage in Ref.~\cite{Bender:1999ek}. In this case, Green's functions of the theory are represented analytically, and therefore, it is straightforward to understand the effect of the background on the interactions that remain valid even in the strongly-coupled regime \cite{Frasca:2015yva}. 
This technique has been recently applied to QCD in Refs.~\cite{Frasca:2021yuu,Frasca:2021mhi,Frasca:2022lwp,Frasca:2022pjf,Chaichian:2018cyv} and to the SM Higgs sector in Ref.~\cite{Frasca:2015wva}, as well as to other types of models over the past two decades in Refs.~\cite{Frasca:2019ysi, Chaichian:2018cyv, Frasca:2017slg, Frasca:2016sky, Frasca:2015yva, Frasca:2015wva, Frasca:2013tma, Frasca:2012ne, Frasca:2009bc, Frasca:2010ce, Frasca:2008tg, Frasca:2009yp, Frasca:2008zp, Frasca:2007uz, Frasca:2006yx, Frasca:2005sx, Frasca:2005mv, Frasca:2005fs}. Recently, some of the authors have employed this technique to study non-perturbative hadronic contributions to the muon anomalous magnetic moment (g-2)$_{\mu}$ \cite{Frasca:2021yuu}, QCD in the non-perturbative regime \cite{Frasca:2021mhi,Frasca:2022lwp,Frasca:2022pjf}, non-perturbative false vacuum decay \cite{Frasca:2022kfy}, as well as to explore the mass gap and confinement in string-inspired infinite-derivative and Lee-Wick theories \cite{Frasca:2020jbe,Frasca:2020ojd,Frasca:2021iip}.

Given the above {\it proviso}, we investigate, in this paper, non-perturbative properties of a simple Yang-Mills theory and show a derivation of the mass gap in the SU(N) gauge sector, condensation of gauge fields and a gap equation for the gauge bosons. 
% Removed after referee's comment.
%This mass gap is then dimensionally transmuted to the electroweak sector and generates the mass for the Higgs boson while avoiding quadratically diverging mass terms in the theory.
%
% Added after referee's comments
We emphasize that our approach does not address the hierarchy problem that, probably, will involve some serious rethinking of fundamental physics.

The paper is organized as follows: in Sec.~\ref{sec2}, we briefly review the non-perturbative technique to understand the strongly-coupled gauge sector  
% Modified after referee's comments
collecting the main equations we need.
% In Sec.~\ref{sec3}, we derive the $\beta$-function for a Yang-Mills theory. 
%
Following 
%
% this result, 
these results,
we study in Sec.~\ref{sec4} the dynamical generation of the EW scale. In Sec.~\ref{sec5}, we study the 
% Removed after referee's comments
%glueball 
mass spectrum of the theory.
%. 
Sec.~\ref{conc} is devoted to conclusions and discussions.  

\medskip

\medskip

\section{Strongly-coupled Yang-Mills: Short Review}
\label{sec2}

%\ag{Short review of the previous work nd introducing the reader to exact Green's function solution and background, etc. Just quote the basic necessary expressions from Ref. \cite{Chaichian:2018cyv}.}

% Added after referee's comments
The aim of this review is to make the paper self-contained presenting already published material and to obtain the main equations we will use in the following for a strongly coupled Yang-Mills theory. Details of the computations are given in the appendixes.

\subsection{Correlation functions}

For a Yang-Mills theory, one has the Lagrangian 
%(to fix the notation we assume $(1,-1,-1,-1)$ for the metric signature)
%
\begin{equation}
\label{eq:L1}
%\label{lagrangian}
{\cal L}={\cal L}_{{inv}}+{\cal L}_{{gf}}+{\cal
L}_{{FP}},
\end{equation}
where ${\cal L}_{{inv}}$ denotes the classical gauge-invariant part, ${\cal L}_{{gf}}$ the gauge-fixing terms, and ${\cal L}_{{FP}}$ the Faddeev-Popov (FP) ghost term characteristic of non-Abelian gauge theories \cite{Smilga:2001ck}:
\begin{eqnarray}
\label{eq:L2} 
{\cal L}_{{inv}}&=&-\frac{1}{4}F_{\mu\nu}\cdot F^{\mu\nu}\,,\cr
%
% Referee's comments 07/02/2020
%{\cal L}_{{gf}}&=&\partial_\mu B\cdot A^\mu+\frac{1}{2}\alpha B\cdot
%B\,,\cr
{\cal L}_{{gf}}&=&-\frac{1}{2\xi}(\partial\cdot A)^2\,,\cr
{\cal L}_{{FP}}&=& -\bar c\cdot\partial_\mu D^\mu c\,,\label{lagr_terms} 
\end{eqnarray}
%
% Added after referee's comments
where $F_{\mu\nu}^a$ is the field tensor (defined below), $A_\mu^a$ are the potentials, ${\bar c}^a,\ c^a$ the Faddeev-Popov ghosts and,  
in the usual notation, $\xi$ is the gauge parameter, and $D_\mu$ represents the covariant derivative whose explicit form is given by
\begin{eqnarray} 
%D_\mu\ \psi&=&(\partial_\mu-igT\cdot A_\mu)\psi\,,\cr
%
D_\mu\ c^a&=&\partial_\mu c^a+g(T^a)^{bc}A_\mu^b c^c\,\label{cov_deriv}
\end{eqnarray}
with $(T^a)^{bc}$ being the gauge group generators and $g$ being the gauge coupling constant. The field strength is given by
\be
F_{\mu\nu}^a=\partial_\mu A_\nu-\partial_\nu A_\mu+gf^{abc}A_\mu^bA_\nu^c \, , 
\ee
where $f^{abc}$ are the structure constants of the gauge group.
% Added after referee's comments
Then, the partition function can be written down as
\be
Z[j,\eta,{\bar\eta}]=\int[dA][d\bar c][dc]e^{-\int d^4x ({\cal L}_{{inv}}+{\cal L}_{{gf}}+{\cal L}_{{FP}}+j\cdot A+\bar\eta\cdot c+{\bar c}\dot\eta)},
\ee
where we have introduced a set of arbitrary bosonic currents $j_\mu^a$ and fermionic sources ${\bar\eta}^a,\ \eta^a$ that are needed to evaluate the correlation functions. Given the partition function, the correlation functions for $n>1$ are defined by
\bea
&&G_{n\mu_1\mu_2\ldots\mu_n}^{a_1a_2\ldots a_n}(x_1,x_2,\ldots,x_n)
=\left.\frac{\delta^n\ln Z}{\delta j^{a_1}_{\mu_1}(x_1)\delta j^{a_2}_{\mu_2}(x_2)\ldots\delta j^{a_n}_{\mu_n}(x_n)}\right|_{j=0}, \nonumber \\
&&K_{n\mu_1\mu_2\ldots\mu_n}^{a_1a_2\ldots a_n}(x_1,x_2,\ldots,x_n)
=\left.\frac{\delta^n\ln Z}{\delta \eta^{a_1}_{\mu_1}(x_1)\delta \eta^{a_2}_{\mu_2}(x_2)\ldots\delta \eta^{a_n}_{\mu_n}(x_n)}\right|_{{\hat\eta},\ \eta=0}, \nonumber \\
&&{\bar K}_{n\mu_1\mu_2\ldots\mu_n}^{a_1a_2\ldots a_n}(x_1,x_2,\ldots,x_n)
=\left.\frac{\delta^n\ln Z}{\delta{\bar\eta}^{a_1}_{\mu_1}(x_1)\delta{\bar\eta}^{a_2}_{\mu_2}(x_2)\ldots\delta{\bar\eta}^{a_n}_{\mu_n}(x_n)}\right|_{{\bar\eta},\ \eta=0}. \nonumber \\
\eea
We have also used occasionally the mixed correlation functions that will be dubbed with the letters $W$, $J$ and $L$ but appear to be 0 in our context limited to $n=2$. These correspond to mixed vertexes in the theory for interactions between ghosts and gluons.

To explore a gauge theory in a non-perturbative regime, we apply the technique with the Dyson-Schwinger equation in differential form. This approach was devised in \cite{Bender:1999ek} and later on studied in several UV-completion scenarios involving string-field and higher-derivative theories \cite{Frasca:2020jbe,Frasca:2020ojd,Frasca:2021iip}. We briefly review the technique in Appendix A for the scalar field. For a Yang-Mills field, we get for the 1P-correlation function
%For the 1P-functions, after performing an average the equations of motion, we get
\be
\Box G_{1\mu}^{(j)a}+gf^{abc}
 \left\langle\partial_\nu{\bar A}^{b}_\mu
  A^{c\nu}\right\rangle+
gf^{abc} \left\langle A^{b\nu}
  (\partial_\mu A^{c}_\nu-\partial_\nu A^{c}_\mu)\right\rangle
g^2f^{abc}f^{cde} 
\left\langle A^{b\nu}  A^{d}_\nu
  A^{e}_\mu\right\rangle
%gf^{abc} \left\langle\left[   \partial_\nu A^{b\nu}
%   A^{c}_\mu\right]\right\rangle-&& \nonumber \\
+gf^{abc}\left\langle \bar{c}^b\partial_\mu c^c\right\rangle=j^a_\mu.
\ee
%and the same for the ghost
Similarly, for the ghost field is
\be
-\Box P_1^{(\eta)a} +gf^{abc}\left\langle\left(A_\mu^c\right)\partial^\mu c^b\right\rangle=\eta^a.
\ee
%For our purposes, we introduced the following 1P-functions
For our needs, we consider the following functions
\bea
\label{eq:defs}
G_{1\mu}^{(j)a}(x)&=&Z^{-1}\langle A_\mu^a(x)\rangle \nonumber \\
P_1^{(\eta)a}(x)&=&Z^{-1}\langle c^a(x)\rangle,
\eea
%The same should hold for ${\bar c}^a$ yielding ${\bar P}_1^{(\eta)a}(x)$. 
and similarly for ${\bar c}^a$ yielding ${\bar P}_1^{(\eta)a}(x)$.
%In order to evaluate these averages we consider the above definitions to be rewritten as
We can evaluate the averages as
\bea
Z[j,\eta,{\bar\eta}]G_{1\mu}^{(j)a}(x)&=&\langle A_\mu^a(x)\rangle, \nonumber \\
Z[j,\eta,{\bar\eta}]P_1^{(\eta)a}(x)&=&\langle c^a(x)\rangle.
\eea
% The apexes $(j)$ and $(\eta)$ remind the explicit dependence on the currents. Let us derive one time with respect to $j(x)$ on the first equation to obtain
We derive one time again with respect to $j(x)$ on the first equation yielding
\be
\label{eq:f1}
Z[j,\eta,{\bar\eta}]G_{2\mu\nu}^{(j)ab}(x,x)+
Z[j,\eta,{\bar\eta}]G_{1\mu}^{(j)a}(x)G_{1\nu}^{(j)b}(x)=
\langle A_\mu^a(x)  A_\nu^b(x)\rangle,
\ee
%We apply the space-time derivative $\partial^\nu$ obtaining
and apply the derivative $\partial^\nu$
\be
Z[j,\eta,{\bar\eta}]\partial^\nu G_{2\mu\nu}^{(j)ab}(x,x)+
Z[j,\eta,{\bar\eta}]\partial^\nu G_{1\mu}^{(j)a}(x)G_{1\nu}^{(j)b}(x)=
\langle\partial^\nu A_\mu^a(x)A_\nu^b(x)\rangle.
\ee
%This step is important as such averages enter into the equation for the 1P-function. 
%We further derive Eqn.(\ref{eq:f1}) with respect to $j^{c\nu}$ to obtain the following:
Then, a further derivation of Eqn.(\ref{eq:f1}) with respect to $j^{c\nu}$ yields
\bea
Z[j,\eta,{\bar\eta}]G_{2\mu\nu}^{(j)ab}(x,x)G_1^{(j)\nu c}(x)
+Z[j,\eta,{\bar\eta}]G_{3\mu\nu}^{(j)abc\nu}(x,x,x)+
\nonumber \\
Z[j,\eta,{\bar\eta}]G_{1\mu}^{(j)a}(x)G_{1\nu}^{(j)b}(x)G_1^{(j)\nu c}(x)+Z[j,\eta,{\bar\eta}]G_{2\mu}^{(j)ac\nu}(x)  G_{1\nu}^{(j)b}+\nonumber \\
Z[j,\eta,{\bar\eta}]G_{2\nu}^{(j)bc\nu}(x)G_{1\mu}^{(j)a}(x)=
\langle A_\mu^a(x)A_\nu^b(x)A^{c\nu}(x)\rangle,
\eea
%and then we need to do the same for the ghost field. From Eqn.(\ref{eq:defs}) we get
and the same to be done on the ghost field. Eqn.(\ref{eq:defs}) yields
\be
\label{eq:P1}
Z[j,\eta,\bar\eta]P_1^{(\eta)a}(x)=\langle c^a(x)\rangle.
\ee
%After deriving with respect to $\partial_\mu$ and then with respect to $\bar\eta$, one gets
The following equation is obtained after deriving with respect to $\partial_\mu$ and then with respect to $\bar\eta$,
\be
Z[j,\eta,{\bar\eta}]{\bar P}_1^{(\eta)b}(x)\partial^\mu P_1^{(\eta)a}(x)+Z[j,\eta,{\bar\eta}]\partial^\mu K_2^{(\eta)ab}(x,x)=
\langle{\bar c}^b\partial^\mu c^a(x)\rangle.
\ee
%We have introduced a new 2P-function which is sdefined as 
A new 2P-function has been introduced and defined as
\be
K_2^{(\eta)ab}(x,y)=\frac{1}{Z[j,\eta,{\bar\eta}]}\frac{\delta P_1^{(\eta)a}(x)}{\delta \eta^b(y)},
\ee
%\ag{Somewhere in the text write in a motivating manner where exactly we wish to reach by doing the steps of the algebra.....}
and the other 2P-functions
\be
J_{2\mu}^{(\eta,j)ab}(x,y)=\frac{1}{Z[j,\eta,{\bar\eta}]}\frac{\delta P_1^{(\eta)a}(x)}{\delta j^{b\mu}(y)}.
\ee
%So, by deriving Eqn.(\ref{eq:P1}) with respect to $j^{b\mu}(x)$, the result is
Then, we derive Eqn.(\ref{eq:P1}) with respect to $j^{b\mu}(x)$ and obtain
\be
Z[j,\eta,{\bar\eta}]G_{1\mu}^{(j)b}(x)\partial^\mu P_1^{(\eta)a}(x)+Z[j,\eta,{\bar\eta}]\partial^\mu J_{2\mu}^{(\eta,j)ab}(x,x)=
\langle A^b_\mu(x)\partial^\mu c^a(x)\rangle.
\ee
%Collecting everything together, one has
Gathering everything together yields
\begin{eqnarray}
\label{eq:G1j}
\Box G_{1\mu}^{(j)a}+gf^{abc}
 \partial^\nu\left[
  G_{2\mu\nu}^{(j)bc}(x,x)+
  G_{1\mu}^{(j)b}(x)  G_{1\nu}^{(j)c}(x)
\right]-&& \nonumber \\
gf^{abc} 
\left[ 
%  A^{b\nu}
%  \partial_\nu A^{c}_\mu\right
  \partial^\nu G_{2\mu\nu}^{(j)bc}(x,x)+
  \partial^\nu G_{1\mu}^{(j)b}(x)  G_{1\nu}^{(j)c}(x)
\right]-
&& \nonumber \\
gf^{abc} 
\left[ 
  \partial_\mu G_{2\nu}^{(j)bc\nu}(x,x)+
  \partial_\mu G_{1\nu}^{(j)b}(x)  G_{1}^{(j)c\nu}(x)
\right]+
&& \nonumber \\
g^2f^{abc}f^{cde} 
\left[
%  A^{b\nu}  A^{d}_\nu
%  A^{e}_\mu
  G_{2\mu\nu}^{(j)bd}(x,x)  G_1^{(j)\nu e}(x)
+  \partial^\nu G_{3\mu\nu}^{(j)bde\nu}(x,x,x)+
\right.
\nonumber \\
  G_{1\mu}^{(j)b}(x)  G_{1\nu}^{(j)d}(x)  G_1^{(j)\nu e}(x)+  G_{2\mu}^{(j)be\nu}(x,x)  G_{1\nu}^{(j)d}(x)+\nonumber \\
\left.
  G_{2\nu}^{(j)de\nu}(x,x)  G_{1\mu}^{(j)b}(x)
\right]-&& \nonumber \\
gf^{abc}  
%\bar{c}^b\partial^\mu\left[(T^a)^{bc}c^c\right]
\left\{
{\bar P}_1^{(\eta)b}(x)  \left[\partial_\mu P_1^{(\eta)c}(x)\right]+\partial_\mu\left[K_2^{(\eta)bc}(x,x)\right]
\right\}
&=& \nonumber \\
  j^a_\mu,&&
\end{eqnarray}
%\ag{Write some expressions, and texts in order to make the idea clear that $M \rightarrow \infty$, we recover the Local theory limit.}
% The equation of the local theory given in \cite{Frasca:2015yva} are now easily obtained by setting the non-locality factor to 1, corresponding to the local limit $M \rightarrow \infty$
%It is possible to recover the local case discussed in \cite{Frasca:2015yva} by setting the non-locality factor to 1, or just taking the limit $M \rightarrow \infty$.
%We see that, in the local limit, we have recovered the Dyson-Schwinger equation for the 1P-function of the Yang-Mills theory \cite{Frasca:2015yva}.
For the ghost field, it is
\be
-\Box P_1^{(\eta)c} 
%+ig\left(  A_\mu^c\right)\partial^\mu c^b
-gf^{abc}G_{1\mu}^{(j)a}(x)\partial^\mu P_1^{(\eta)b}(x)-gf^{abc}\partial^\mu J_{2\mu}^{(\eta,j)ab}(x,x)
=\eta^c.
\ee
%Fixing $(T^c)^{ab}=f^{abc}$, the fundamental representation, one has
%\begin{eqnarray}
%\Box G_{1\mu}^{(j)a}+gf^{abc}
% \partial^\nu\left[
%  G_{2\mu\nu}^{(j)bc}(x,x)+
%  G_{1\mu}^{(j)b}(x)  G_{1\nu}^{(j)c}(x)
%\right]-&& \nonumber \\
%g^2f^{abc}f^{cde} 
%\left[
%%  A^{b\nu}  A^{d}_\nu
%%  A^{e}_\mu
%  G_{2\mu\nu}^{(j)bd}(x,x)  G_1^{(j)\nu e}(x)
%+  \partial^\nu G_{3\mu\nu}^{(j)bde\nu}(x,x,x)+
%\right.
%\nonumber \\
%  G_{1\mu}^{(j)b}(x)  G_{1\nu}^{(j)d}(x)  G_1^{(j)\nu e}(x)+  G_{2\mu}^{(j)be\nu}(x,x)  G_{1\nu}^{(j)d}(x)+\nonumber \\
%\left.
%  G_{2\nu}^{(j)de\nu}(x,x)  G_{1\mu}^{(j)b}(x)
%\right]-&& \nonumber \\
%gf^{abc} 
%\left[ 
%%  A^{b\nu}
%%  \partial_\nu A^{c}_\mu\right
%  \partial^\nu G_{2\mu\nu}^{(j)bc}(x,x)+
%  \partial^\nu G_{1\mu}^{(j)b}(x)  G_{1\nu}^{(j)c}(x)
%\right]-
%&& \nonumber \\
%gf^{abc} 
%\left[ 
%  A^{b\nu}
%  \partial_\nu A^{c}_\mu\right
%  G_{1\mu}^{(j)b}(x)  \partial^\nu G_{1\nu}^{(j)c}(x)\right]-
%&& \nonumber \\
%ig
%\bar{c}^b\partial^\mu\left[(T^a)^{bc}c^c\right]
%f^{abc}  \left\{
%{\bar P}_1^{(\eta)b}(x)  \left[\partial_\mu P_1^{(\eta)c}(x)\right]+\partial_\mu K_2^{(\eta)bc}(x,x)
%\right\}
%&=& \nonumber \\
%  j^a_\mu,&&
%\end{eqnarray}
%and
%\be
%-\Box P_1^{(\eta)c} 
%+ig\left(  A_\mu^c\right)\partial^\mu c^b
%-ig  f^{abc}G_{1\mu}^{(j)a}(x)\partial^\mu P_1^{(\eta)b}(x)-igf^{abc}\partial^\mu J_{2\mu}^{(\eta,j)ab}(x,x)
%=  \eta^c.
%\ee
% After setting all the current to zero, the Dyson-Schwinger equations for the 1P-functions are obtained in the form as given in the main text.
The final step is given by setting all the currents to zero.

%For the 2P-functions, let us consider firstly Eqn.(\ref{eq:G1j}) and 
%From Eqn.(\ref{eq:G1j}), we derive it with respect to $j^{\lambda h}(y)$ and get
From Eqn.(\ref{eq:G1j}), we derive it with respect to $j^{\lambda h}(y)$ and get
% Changed the metric from \eta to g: Too much \eta around
\begin{eqnarray}
\label{eq:G2j}
\Box G_{2\mu\lambda}^{(j)ah}(x,y)+gf^{abc}
 \partial^\nu\left[
  G_{3\mu\nu\lambda}^{(j)bch}(x,x,y)+
  G_{2\mu\lambda}^{(j)bh}(x,y)
  G_{1\nu}^{(j)c}(x)+G_{1\mu}^{(j)b}(x)
  G_{2\nu\lambda}^{(j)ch}(x)
\right]-&& \nonumber \\
gf^{abc} 
\left[ 
%  A^{b\nu}
%  \partial_\nu A^{c}_\mu\right
  \partial^\nu G_{2\mu\nu\lambda}^{(j)bch}(x,x,y)+
  \partial^\nu G_{2\mu\lambda}^{(j)bh}(x,y)  G_{1\nu}^{(j)c}(x)+
  \partial^\nu G_{1\mu}^{(j)b}(x)  G_{2\nu\lambda}^{(j)ch}(x,y)\right]-&& \nonumber \\
gf^{abc} 
\left[
  \partial_\mu G_{3\nu\lambda}^{(j)bch\nu}(x,x,y)+
  \partial_\mu G_{2\nu\lambda}^{(j)bh}(x,y)  G_{1}^{(j)c\nu}(x)+
  \partial_\mu G_{1\nu}^{(j)b}(x)  G_{2\lambda}^{(j)ch\nu}(x,y)\right]+&& \nonumber \\
g^2f^{abc}f^{cde} 
\left[
%  A^{b\nu}  A^{d}_\nu
%  A^{e}_\mu
  G_{3\mu\nu\lambda}^{(j)bdh}(x,x,y)  G_1^{(j)\nu e}(x)+
  G_{2\mu\nu}^{(j)bd}(x,x)  G_{2\lambda}^{(j)\nu eh}(x,y)
+  \partial^\nu G_{4\mu\nu\lambda}^{(j)bdeh\nu}(x,x,x,y)+\right.&& \nonumber \\
  G_{2\mu\lambda}^{(j)bh}(x,y)  G_{1\nu}^{(j)d}(x)  G_1^{(j)\nu
e}(x)+
  G_{1\mu}^{(j)b}(x)G_{2\nu\lambda}^{(j)dh}(x,y)  G_1^{(j)\nu
e}(x)+&&
\nonumber \\
  G_{1\mu}^{(j)b}(x)G_{1\nu}^{(j)d}(x)G_{2\lambda}^{(j)\nu
eh}(x,y)+
  G_{3\mu\lambda}^{(j)beh\nu}(x,x,y)G_{1\nu}^{(j)d}(x)+
  G_{2\mu}^{(j)be\nu}(x,x)G_{2\nu\lambda}^{(j)dh}(x,y)+&&
\nonumber \\
\left.
  G_{3\nu\lambda}^{(j)deh\nu}(x,x,y)  G_{1\mu}^{(j)b}(x)
+  G_{2\nu}^{(j)de\nu}(x,x)  G_{2\mu\lambda}^{(j)bh}(x,y)
\right]-
gf^{abc}  
%\bar{c}^b\partial^\mu\left[(T^a)^{bc}c^c\right]
\left\{
{\bar J}_{2\lambda}^{(\eta,j)bh}(x,y)  \left[\partial_\mu P_1^{(\eta)c}(x)\right]\right.+&& \nonumber \\
\left.{\bar P}_1^{(\eta)b}(x)  \left[\partial_\mu J_{2\lambda}^{(\eta)ch}(x,y)\right]
+\partial_\mu\left[W_{3\lambda}^{(\eta,j)bch}(x,x,y)\right]
\right\}=\delta^{ah}g_{\mu\lambda}\delta^4(x-y),&&
\end{eqnarray}
after the introduction of the 3P-function
\be
W_{3\lambda}^{(\eta,j)abc}(x,y,z)=Z^{-1}\frac{\delta K_2^{(\eta)ab}(x,y)}{\delta j^{\lambda c}(z)}.
\ee
%Similarly, starting right from the 1P-function for the ghost and deriving it with respect to $\eta^{h}(y)$, we get
Taking the 1P-function for the ghost, we derive it with respect to $\eta^{h}(y)$ and obtain
\bea
&-\Box K_2^{(\eta)ch}(x,y)
%+ig\left(  A_\mu^c\right)\partial^\mu c^b
-igf^{abc}L_{2\mu}^{(\eta,j)ah}(x,y)\partial^\mu P_1^{(\eta)b}(x)\nonumber \\
&-igf^{abc}  G_{1\mu}^{(j)a}(x)\partial^\mu K_2^{(\eta)bh}(x,y)
-igf^{abc}\partial^\mu W_{3\mu}^{(\eta,j)abh}(x,x,y) \nonumber \\
&=  \delta^{ch}\delta^4(x-y).
\eea
%\ag{Write some expressions, and texts in order to make the idea clear these are some interesting results and where we are heading next using these.}
We have introduced the 2P-function
\be
L_{2\mu}^{(\eta,j)ab}(x,y)=\frac{\delta G_1^{(j)a}(x)}{\delta \eta^b(y)}.
\ee
%Deriving with respect to $j^{h\nu}(y)$, one has the equation for $J_2$ in the form
We derive with respect to $j^{h\nu}(y)$ and this yields the equation for $J_2$ as
\bea
&-\Box J_2^{(\eta)ch\nu}(x,y) 
%+ig\left(  A_\mu^c\right)\partial^\mu c^b
-igf^{abc}  G_{2\mu\nu}^{(j)ah}(x,y)\partial^\mu P_1^{(\eta)b}(x) \nonumber \\
&-igf^{abc}  G_{1\mu}^{(j)a}(x)\partial^\mu J_2^{(\eta,j)bh\nu}(x,y) \nonumber \\
&-igf^{abc}\partial^\mu J_{3\mu}^{(\eta,j)abh}(x,x,y)=0,
\eea
with the introduction of the 3P-function
\be
J_{3\mu}^{(\eta,j)abc}(x,y,z)=\frac{\delta J_{2\mu}^{(\eta,j)ab}(x,y)}{\delta j^{c\mu}(z)}.
\ee
%We can easily now recover the equations in the main text after setting all the currents to zero.
%Again, we repeat the final step setting all the currents to zero. This produces the equations given in the main text.
% Added after referee's comments
This set of equations is amenable to an exact solution \cite{Frasca:2015yva} that is of a Gaussian kind \cite{Frasca:2023uaw} with all higher order correlation functions obtained by 1P- and 2P-correlation functions, provided the following conditions are satisfied:
\begin{enumerate}
    \item The mapping theorem, presented in \cite{Frasca:2009yp,Frasca:2007uz}, is used to solve for $G_1$ and the remaining equations.
    \item All the nP-correlation functions for $n>2$ are 0 when two or more independent variables coincide. This arises from the property of the support of $G_2$ entering in the solutions in the same way as the product $\theta(x)\theta(-x)$ under integration, where $\theta(x)$ is the Heaviside step function.
\end{enumerate}
One can verify that the solution obtained in this way trivially satisfies the Slavnov-Taylor identities \cite{Pascual:1984zb} at the non-perturbative level.

In the following, we will use the solutions of the 2P-correlation functions. 
% Removed after referee's comment
%These are given by, 
The propagator in the Landau gauge is given by 
% Removed after referee's comment
% for the Faddeev-Popov ghost field as
% \be
% K_2(p)=-\frac{1}{p^2+i\epsilon}
% \ee
%
%\ag{We need to define K2.}
%and the same for the ghost field to be
%and for the gauge field, in the same gauge choice, one has
% Added after referee's comments
(see Appendix C for a derivation)
\be
\label{eq:G2sol}
G_2(p)=\frac{\pi^3}{4K^3(i)}
	\sum_{n=0}^\infty\frac{e^{-(n+\frac{1}{2})\pi}}{1+e^{-(2n+1)\pi}}(2n+1)^2\frac{1}{p^2-m_n^2+i\epsilon}.
\ee
%for the gauge field, with the mass spectrum \ag{whose mass psectrum ?}
We observe that the gauge field 2P-function is in agreement with the K\"allen-Lehman representation with a spectrum of particles
% Added after referee's comments
% Removed after referee's comments
% the glueball spectrum (see below),
%
%
given by
\be
\label{eq:glue}
m_n=(2n+1)\frac{\pi}{2K(i)}\left(\frac{Ng^2}{2}\right)^\frac{1}{4}\mu,
\ee
where $K(i)$ is the complete elliptical integral of the first kind and $\mu$ is one of the integration constants of the theory \cite{Frasca:2015yva}. At this stage we have neglected the quantum corrections inducing a gap equation for the gauge field \cite{Frasca:2017slg}, so that the gauge field propagator is a good approximation to the exact one provided that the mass shift induced by quantum fluctuations is small. This is what one sees for SU(N) Yang-Mills theory \cite{Frasca:2017slg}. In the Landau gauge, the gauge field 2P-function can be written down in the form
\be
D_{\mu\nu}^{ab}(p)=\delta_{ab}\left(\eta_{\mu\nu}-\frac{p_\mu p_\nu}{p^2}\right)G_2(p),
\ee
where $\eta_{\mu\nu}$ is the Minkowski metric tensor.
% Added after referee's comment
% Removed after referee's comments
% To prove that eq.(\ref{eq:glue}) is indeed the spectrum of glueballs,
% we consider the correlation function for the
% scalar glueballs~\cite{Narison:2002woh,Narison:2021xhc}
% \begin{equation}
% \label{eq:4corr}
% {\cal O}(x)=\langle F^{a\mu\nu}(x)F^a_{\mu\nu}(x)
%   F^{b\rho\eta}(0)F^b_{\rho\eta}(0)\rangle.
% \end{equation}
% The poles of this correlation function are the physical glueballs. 
% With the method discussed above and presented firstly in Ref.~\cite{Frasca:2015yva}, one can see that,
% according to Ref.~\cite{Windisch:2012sz}, the four-point correlator (\ref{eq:4corr}) can be reduced to integrated products of
% 1P- and 2P-correlation functions. As the 1P-correlation function has no poles but just
% zeros, the poles of the glueball four-point correlator (\ref{eq:4corr}), are given by the poles
% of the two-point correlator given by eq.(\ref{eq:glue}). Therefore, these poles represent true colorless
% glueball states. 
%
%

\medskip

% Changed after referee's comments
% \section{$\beta$-function for Strongly-Coupled Yang-Mills Theory}
% \label{sec3}
\subsection{$\beta$-function for Strongly-Coupled Yang-Mills Theory}
%

%\ag{Please add a brief review discussion on the $\beta-$ function for our model using your technique.}
%\section{Dynamical Generation Higgs Mass}
Following 
% Modified after referee's comment
%an exact solution 
an exact solution of a truncated set of the background equations of motion
to the Yang-Mills fields using elliptical integral equation,
$\beta-$function for Yang-Mills theory has been obtained recently, in an almost exact form, in Ref. \cite{Chaichian:2018cyv} and then later on extended onto various other extensions of QFT in the context of UV-completion \cite{Frasca:2020jbe,Frasca:2020ojd,Frasca:2021iip}. 
% Added after referee's comments
We present this argument here in order to make the paper self-contained as much as possible.
The idea is to apply the Kugo-Ojima criterion for confinement starting from the exact solutions obtained for the Dyson-Schwinger equations in \cite{Frasca:2015yva}. Kugo-Ojima criterion requires \cite{Kugo:1977zq,Kugo:1979gm}
\begin{equation}
\label{eq:KOc}
\int d^dxe^{ipx}\langle D_\mu\bar{c}^{a} (x),D_\nu c^{b} (y) \rangle=\delta^{ab}
\left(\delta_{\mu\nu} - \frac{p_{\mu} p_{\nu}}{p^2-i\epsilon}\right) u(p^2)-\delta^{ab}\frac{p_{\mu} p_{\nu}}{p^2-i\epsilon},
\end{equation}
%\ag{Let us give the expression for D.}
%\ag{Let us give the expression for u.}
where $D_\mu=\partial_\mu-igT^aA_\mu^a$ is the covariant derivative with $T^a$ the group generators and $A_\mu^a$ the gauge potential and the $u$-function derived below. 
% Added after referee's comment
We expect that no massless particle propagates in our theory and we have
the no-pole condition that
%, $C=0$ for the other approach in eq. (\ref{eq:KO}), 
yields here
\begin{equation} 
1+u(p^2=0)=0,
\end{equation}
which is the Kugo--Ojima condition for confinement
% Added after referee's comment
\cite{Kugo:1977zq,Kugo:1979gm}\footnote{Eq.(\ref{eq:KOc}) and the confinement condition are derived through the anti-commuting charges of BRST symmetry (see \cite{Chaichian:2018cyv} for a simplified derivation).}. In this equation, $D_\mu$ is the covariant derivative and $c,\ {\bar c}$ are the ghost fields. Therefore, BRST invariance plays an essential role in this argument.

% The gluon propagator is given by \cite{Frasca:2015yva}
% \begin{equation}
%   \Delta(p)=\frac{\pi^3}{4K^3(-1)}
% 	\sum_{n=0}^\infty\frac{e^{-(n+\frac{1}{2})\pi}}{1+e^{-(2n+1)\pi}}(2n+1)^2\frac{1}{p^2-m_n^2+i\epsilon},
% \end{equation}
% with $K(-1)$ being an elliptic integral that yields the numerical constant $1.3110287771460598\ldots$ while for the mass spectrum one has
% \begin{equation}
% \label{eq:spec}
%   m_n=(2n+1)\frac{\pi}{2K(-1)}\left(\frac{Ng^2}{2}\right)^\frac{1}{4}\sigma^\frac{1}{2}.
% \end{equation}
% This is an approximation as we are neglecting a mass renormalization shift. The constant $\sigma$ arises by integration of the theory and has the dimension of a mass. For the ghost propagator, it decouples from the theory and we can use for it a free massless propagator. 
Thus, the $u$ function will be given by \cite{Chaichian:2018cyv}
\begin{equation}
  u(p^2)=-\frac{(N^2-1)^2}{2N}g^2\int\frac{d^4p'}{(2\pi)^4}\frac{1}{|p-p'|^2}G_2(p').
\end{equation}
Then, we have to evaluate the integral
\begin{eqnarray}
  u(0)&=&-\frac{(N^2-1)^2}{2N}g^2\int\frac{d^4p}{(2\pi)^4}\frac{1}{p^2}\sum_{n=0}^\infty B_n\frac{1}{p^2+m_n^2}\cr
&=&
	-\frac{(N^2-1)^2}{2N}g^2\sum_{n=0}^\infty\frac{B_n}{m_n^2}
	\int\frac{d^4p}{(2\pi)^4}\left(\frac{1}{p^2}-\frac{1}{p^2+m_n^2}\right),
\end{eqnarray}
with $B_n=\frac{\pi^3}{4K^3(-1)}\frac{e^{-(n+\frac{1}{2})\pi}}{1+e^{-(2n+1)\pi}}(2n+1)^2$. This integral diverges and needs to be renormalized.
% Added after referee's comments
We get after dimensional regularization
\begin{equation}
\label{eq:u0}
  u(0)=\frac{(N^2-1)^2}{2N}\frac{\alpha}{4\pi}\left[-1+\gamma+\sum_{n=0}^\infty B_n\ln\left(\frac{m_n^2}{4\pi\mu^2}\right)\right],
\end{equation}
where use has been made of the identities $\sum_{n=0}^\infty B_n=1$ and $\alpha=g^2/4\pi$. Finally, using the Kugo-Ojima criterion $u(0)=-1$, we get an equation for the running coupling
\begin{equation}
\label{eq:rc}
    \frac{(N^2-1)^2}{2N}\frac{\alpha(\mu^2)}{4\pi}\left[-1+\gamma+\sum_{n=0}^\infty B_n\ln\left(\frac{m_n^2}{4\pi\mu^2}\right)\right]=-1.
\end{equation}
It should be emphasized that also $m_n$ depends on $\alpha$ and this is a non-linear equation to be solved. Indeed, it can be cast into a differential form yielding
%this computation will provide 
for the $\beta-$function for a SU(N) gauge theory
% Removed after referee's comments
%, using Kugo-Ojima criterion,
%
\begin{equation}
\label{eq:CS}
   \frac{d\alpha}{dl}=-\beta_0\frac{\alpha^2}{1-\frac{1}{2}\beta_0\alpha},
\end{equation}
%This is almost exact as we have neglected the mass shift in the spectrum assuming it being small. 
with $\beta_0=(N^2-1)^2/8\pi N$. We have set $l=\ln(\mu^2/\sigma_0)$ as an independent variable. Such a $\beta$-function is almost exact because in the derivation we neglected a correction to the propagator arising from quantum fluctuations (see eq.~(\ref{eq:dm2}) in appendix C).
% Added after referee's comments
The equation we obtained
has the same functional form of the exact $\beta$-function obtained in supersymmetric Yang-Mills theory \cite{Novikov:1983uc,Shifman:1986zi} and the corresponding one, inspired by the supersymmetric solution, for pure Yang-Mills theory \cite{Ryttov:2007cx}.
%
% Commented out due to referee's comments
% % Added after referee's comments on 27-4-2024
% In the asymptotic freedom regime, the agreement with the perturbative result at the leading order is fairly good. We get, for $N=3$, $\alpha_s=(12\pi/32)\ln^{-1}(\mu^2/\sigma_0)$ to be compared with the perturbative result $\alpha_s=(12\pi/33)\ln^{-1}(\mu^2/\sigma_0)$. Thus, the agreement with experimental data ($\alpha_s(M_Z)$ with $M_Z$ the mass of the Z boson) is excellent as well. Besides, the limit $N\rightarrow\infty$ recovers the strong coupling limit yielding $\beta=2\alpha_s+32\pi/N^3+O(N^{-4})$, that is similar to take the limit $\alpha_s\rightarrow\infty$ in eq.(\ref{eq:CS}). %Finally,
% %
% %consistently with the Kugo-Ojima criterion, this renormalization group equation entails a finite infrared fixed point.
% % Added after referee's comments In second round
% The agreement can also be extended to 1-loop, for the sake of completeness. We have to solve the following equation for $\alpha_s\rightarrow 0$
% \begin{equation}
% \label{eq:CS1}
%    \frac{d\alpha_s}{dl}=-\beta_0\alpha_s^2-\frac{1}{2}\beta_0^2\alpha_s^3+O(\alpha_s^4).
% \end{equation}
% We get, in the limit $l\rightarrow\infty$
% \begin{equation}
%     \alpha_s(l)=\frac{3\pi}{8l}-\frac{3\pi}{16}\frac{\ln l}{l^2}+O(1/l^3).
% \end{equation}
%
% Removed after referee's comments
%Perturbation theory yields $102/121\approx 0.84$ while we get $0.5$ that is a good agreement taking into account the involved approximations.
%

\section{Generating the Electroweak Scale Dynamically}
\label{sec4}

\subsection{Higgs coupled to a gauge field}

%\ag{We can add positive mass term. As long as negative mass dominates positive one, we will have EWSB which is dynamical symmetry breaking.}

\begin{figure}[H]
\centering
\includegraphics[height=7cm,width=8cm]{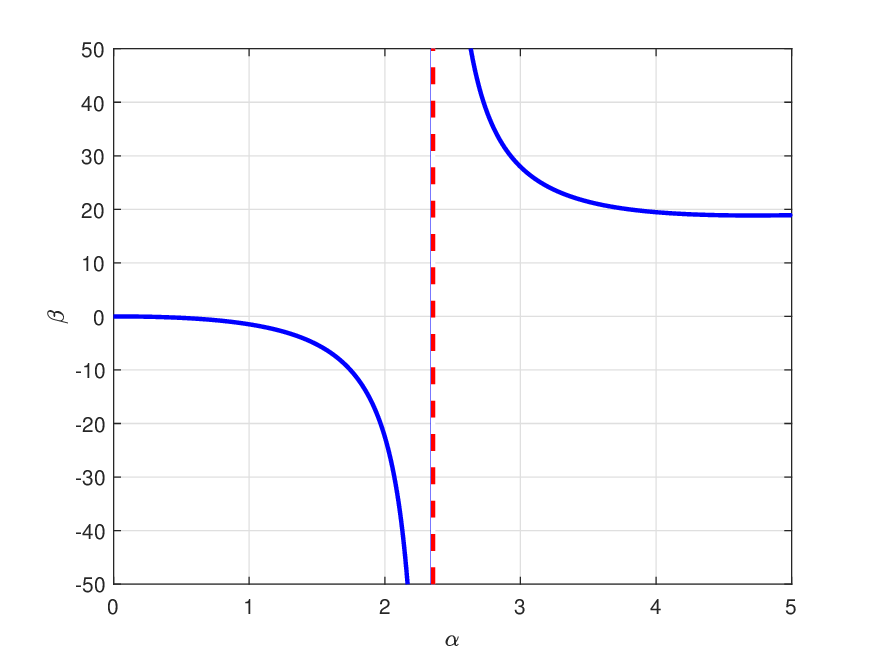}
\caption{\it $\beta$-function for Yang-Mills theory given for SU(3).}
\label{fig:beta}
\end{figure}

\begin{figure}[H]
\centering
\includegraphics[height=7cm,width=8cm]{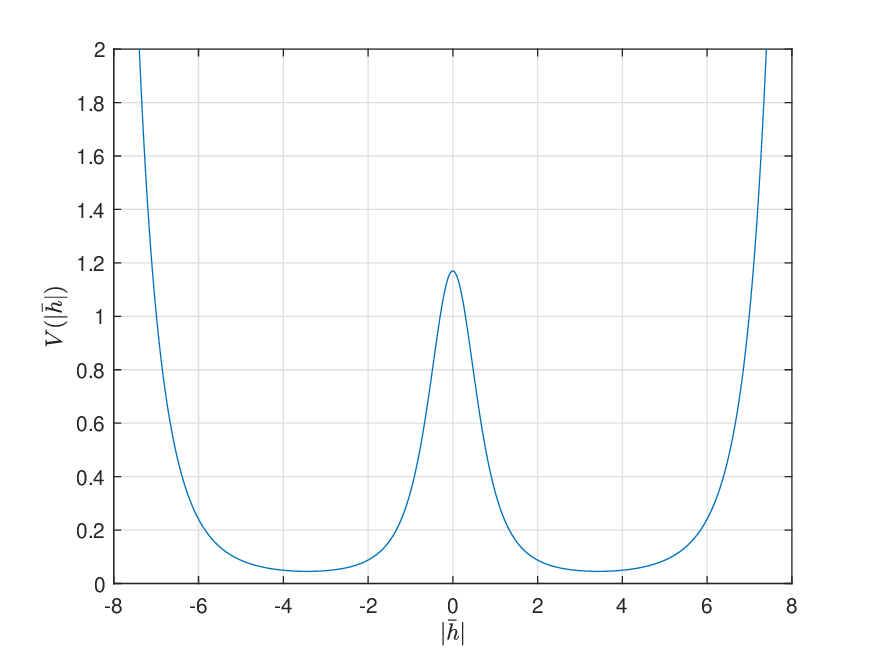}
\caption{\it General behavior of the effective potential obtained from eq.(\ref{eq:Vac}).
%and the quartic potential in the Higgs Lagrangian. 
%\ag{We need to show the plot in terms of model parameters.}
In this case we have normalized $V(|{\bar h}|)$ to $V_0$ to make the potential dimensionless.
%\ag{Let us complete....}.
%\ag{let us mentin what values we used...}
}
\label{fig:pote}
\end{figure}

Let us consider the following Higgs Lagrangian \cite{Chun:2019box}
\be
L=\partial^\mu h^\dagger\partial_\mu h-\lambda(h^\dagger h)^2-\frac{1}{4}\left(1+\frac{h^\dagger h}{M^2}\frac{\beta}{2g}\right)G^2,
\ee
%\ag{Where.....h is, G is....}
%\ag{Let us put in the standard model form with covariant derivatives...}
%\ag{Is this equation OK ?}
where $h$ is a Higgs field
%Modified after referee's comment
and a singlet representation of the SU(N) group,
$\beta$ the beta function we introduced above, $M$ a mass scale, and $G_{\mu\nu}^{a}=\partial_\mu A^a_\nu-\partial_\nu A^a_\mu+gf^{abc}A^b_\mu A^c_\nu$ the field tensor.
Here, the mass term is not present. However, it is also possible to consider a positive mass term $(h^\dagger h)$. As long as the negative mass term dominates the EWSB dynamics via non-perturbative physics, what we describe below remains the same. 
% Added after referee's comments
% We avoid to consider a charged $h$ field because we are interested to observe dimensional transmutation after confinement for the gauge field. This limits our interaction term to a sixth-order operator. 
% Further referee's comments
The field $h$ is not charged to avoid terms like $|h^2|A_\mu^aA^{a\mu}$ and we avoid to break explicitly the gauge invariance. Besides, we work in a confining phase of the gauge field where a renormalization group analysis, as per a UV-regime, cannot apply giving rise to such kind of breaking terms. Then,
the idea is to use the mean-field approximation on the equation of motion
\be
\label{eq:eom}
\Box h=-\lambda (h^\dagger h)h-\frac{h}{M^2}\frac{\beta}{2g}G^2
\ee
and to evaluate the gluon condensate $\langle G^2\rangle$. 
% Removed after referee's comments
%We already did this in \cite{Frasca:2008bz} but we repeat the computation here.
% Added after referee's comments
Our aim is to show how an evaluation of the gluon condensate can be obtained directly from Yang-Mills theory with our approach.
The gluon condensate can be computed starting from
%We use
%
the following solutions for the gauge field \cite{Frasca:2015yva}
% Added after referee's comments
(for an extended explanation of the constants $\eta_\mu^a$ and the scalar field $\phi$ see Appendix B and C)
\be
A_\mu^a(x)=\eta_\mu^a\phi(x).
\ee
% Added after referee's comments on 27-4-2024
This is an application of a mapping theorem \cite{Frasca:2009yp} providing a set of exact solutions for the gauge field starting from a $\phi^4$ theory. For our aims, $\phi$ has just a mathematical meaning here.
%
% Added after referee's comments in second round
It should be emphasized that in some gauge such a theorem just warrants asymptotic and not an exact solution in the limit of the coupling running to infinity. This would not change our argument if we would change our gauge's choice.
%
%\ag{Plot of the beta function with some parameter choice....}
Therefore, one has for SU(N) (see Appendix B)
\be \label{eq:condensate-1}
%\langle G^2\rangle=N(N^2-1)\langle(\partial\phi)^2\rangle-4\pi N(N^2-1)\alpha_s\langle\phi^4\rangle.
\langle G^2\rangle=2(D-2)(N^2-1)\langle(\partial\phi)^2\rangle+4\pi N(N^2-1)\alpha_s\langle\phi^4\rangle.
\ee
% Commented out after referee's comments
%One has \cite{Frasca:2015yva}
To evaluate such a condensate we have to compute ($p_E$ is the Euclidean momentum)
% Modified after referee's comments
\bea
\langle(\partial\phi)^2\rangle&=&-\int\frac{d^4p_E}{(2\pi)^4}p_E^2G_2(p_E), \nonumber \\
%\langle\phi^4\rangle &=& -3\left(\int\frac{d^4p}{(2\pi)^4}G_2(p)\right)^2,
\eea
% where \cite{Frasca:2015yva}
% Changed after referee's comments
where $G_2(p_E)$ is given in eq.(\ref{eq:G2sol}).
% where (see Appendix C for a derivation)
% %
% \begin{equation}
%    G_2(p)=\frac{\pi^3}{4K^3(-1)}
% 	\sum_{n=0}^\infty\frac{e^{-(n+\frac{1}{2})\pi}}{1+e^{-(2n+1)\pi}}(2n+1)^2\frac{1}{p^2-m_n^2+i\epsilon}
% \end{equation}
% $K(-1)$ being the elliptical integral of the first kind. This holds given the mass spectrum
% \begin{equation}
%    m_n=(2n+1)\frac{\pi}{2K(-1)}\left(\frac{Ng^2}{2}\right)^\frac{1}{4}\Lambda,
% \end{equation}
% with $\Lambda$ being an integration constant that we will use as a natural cut-off for the theory. 
%
Using this propagator, we are neglecting the effect of mass renormalization on the spectrum of the theory that we consider negligibly small \cite{Frasca:2017slg}. 
% Removed after referee's comment
% In order to evaluate these integrals, we introduce the string tension (a measurable quantity in strong interaction experiments)
% \be
% \sigma=\sqrt{2\pi N\alpha}\Lambda^2.
% \ee
% At low energies, one has $\sigma\approx(0.44\ \text{GeV})^2$ for the strong interaction. 
%
Finally, we evaluate the integrals by moving the integration path to complex plane changing the metric to Euclidean. This will yield, after we introduce a cut-off $\Lambda$,
% Added after referee's comments
(see Appendix C for a derivation)
%
% Recomputed and fixed after referee's comments
% \bea
% \langle(\partial\phi)^2\rangle&=&\frac{1}{8\pi^2}\frac{\Lambda^4}{4}, \nonumber \\
% \langle\phi^4\rangle &=& -3\left(\frac{1}{8\pi^2}\frac{\Lambda^2}{2}\right)^2.
% \eea
\be
\langle(\partial\phi)^2\rangle=\frac{1}{8\pi^2}\frac{\Lambda^4}{4}.
\ee
This yields the value for the condensate
% \bea \label{eqn:condensate-2}
% \langle\frac{\alpha_s}{\pi}G^2\rangle&=&N(N^2-1)\frac{1}{8\pi^2}\frac{\Lambda^4}{4}+N(N^2-1)\alpha_s\frac{3}{16\pi^3}
% \frac{\Lambda^4}{4} \nonumber \\
% &=&\frac{N(N^2-1)}{32\pi^3}\alpha_s
% \left(1+\frac{3}{2\pi}\alpha_s\right)\Lambda^4.
% \eea
\be \label{eqn:condensate-2}
\langle\frac{\alpha}{\pi}G^2\rangle=\alpha\frac{N^2-1}{8\pi^3}\Lambda^4
\ee
% Added after referee's comments
The $\beta$-function is given by
% changed after referee's comments
%\cite{Chaichian:2018cyv} 
eq.~(\ref{eq:CS}) and can be written as  \cite{Chaichian:2018cyv}
\be \label{eqn:running}
\beta(\alpha)=-\beta_0\frac{\alpha^2}{1-\frac{1}{2}\beta_0\alpha},
\ee
%and is 
% Added after referee's comment
with $\beta_0=(N^2-1)^2/8\pi N$, whose plot is shown in Fig.~\ref{fig:beta}.
% Removed after erferee's comment
% For the running coupling, we can use the result given in \cite{Chaichian:2018cyv}. One has (see Fig.\ref{fig:beta})
% with $\beta_0=(N^2-1)^2/8\pi N$. The minus sign in the $\beta$-function grants a proper sign for the mass term 
% in the equation of motion (\ref{eq:eom}).
%
We can introduce an effective potential by noting that the gauge term in the Lagrangian can have the form,
\be
L_g=-\frac{1}{4}\left(1+\frac{h^\dagger h}{M^2}\frac{\beta}{2g}\right)G^2
=-\frac{1}{4g_s^2}{\tilde G}^2,
\ee
% Changed after referee's comments
with ${\tilde G}=G/g$ 
%and the limit $g\gg 1$. 
and having set
\be
\frac{1}{g_s^2}=\frac{1}{g^2}\left(1+\frac{h^\dagger h}{M^2}\frac{\beta}{2g}\right).
\ee
% Added after referee's comment
Here, $g_s$ should be understood as an effective coupling due to its dependence on $h$. This definition is needed to complete the mathematical steps to derive the potential for $h$ itself.
We are assuming an infrared fixed point at $\alpha_s=\pi$. From the $\beta$ function in eq.(\ref{eqn:running}), one gets the general solution
\be
\alpha(l)=-\frac{2}{\beta_0}\frac{1}{W\left(-Ce^{-2l}\frac{2}{\beta_0}\right)} \, ,
\ee
where $C$ is an integration constant, $W$ is the Lambert function 
% Added after refree's comments
that solves the equation $z=W(z)e^{W(z)}$ and $l=\ln(\Lambda/M)$ the renormalization group scaling with the cut-off $\Lambda$ normalized to the mass $M$. 
%
%This solution displays a finite value of $\alpha$ for $l$ going toward negative values as, beyond a certain value of the argument, the Lambert function becomes imaginary reaching nonphysical values. 
%Therefore, 
%We assume to have fixed the constant $C$ in such a way to have for the running coupling the value $\pi$ in the infrared. This means that, from the Lagrangian, one has
Conventionally, we consider the value of $\beta$ for $\alpha=\pi$ and we notice that this must be a fixed point for the renormalization group of the theory. This gives
\be
\label{eq:alphah}
\frac{1}{\alpha_s}=\frac{1}{\alpha}\left(1+\frac{|h|^2}{M^2}\frac{\beta(\pi)}{4\sqrt{\pi\alpha}}\right)=
\frac{1}{\pi}\left(1+\frac{|h|^2}{M^2}\frac{\beta(\pi)}{4\pi}\right),
\ee
%where use has been made of the fixed point at $\alpha=\pi$ and $\beta(\pi)$ is the $\beta$ function evaluated at that value of the coupling. 
% Added after referee's comments
where $\alpha_s=g_s^2/4\pi$ and similarly for $\alpha$ and $g$.
We can invert our solution for the running coupling to get
% Added after referee's comments
the cut-off $\Lambda$ at the value $\alpha=\pi$ as
%
% Verified after referee's comments
% \be
% \Lambda =M\frac{1}{C^\frac{1}{4}\pi}\frac{1}{\left(1+\frac{|h|^2}{M^2}\frac{\beta(\pi)}{4\pi}\right)^\frac{1}{4}}\exp\left(\frac{1}{2\beta_0}\left(1+\frac{|h|^2}{M^2}\frac{\beta(\pi)}{4\pi}\right)\right).
% \ee
\be
\Lambda =M(\alpha_s C)^\frac{1}{2}\exp\left(\frac{1}{\beta_0\alpha_s}\right)
= M(\pi C)^\frac{1}{2}\frac{1}{\left(1+\frac{|h|^2}{M^2}\frac{\beta(\pi)}{4\pi}\right)^\frac{1}{2}}\exp\left(\frac{1}{\beta_0\pi}\left(1+\frac{|h|^2}{M^2}\frac{\beta(\pi)}{4\pi}\right)\right).
\ee
Therefore, the potential will become
% Added after referee's comments
\cite{Pasechnik:2016sbh}
%
% Verified after referee's comments
% \bea \label{eq:Vac}
% V_{\rm vac} &=& \frac{1}{4}\langle T_\mu^\mu\rangle = \langle\frac{\beta}{8g_s}{\tilde G}^2\rangle =
% \frac{5}{2}\frac{N(N^2-1)}{128\pi^4}\Lambda^4  \nonumber \\
% &=&\frac{5}{2}\frac{N(N^2-1)}{128\pi^8}\frac{M^4}{C}\frac{1}{1+\frac{|h|^2}{M^2}\frac{\beta(\pi)}{4\pi}}\exp\left(\frac{2}{\beta_0}\left(1+\frac{|h|^2}{M^2}\frac{\beta(\pi)}{4\pi}\right)\right).
% \eea
\bea \label{eq:Vac}
V_{\rm vac} &=& \frac{1}{4}\langle T_\mu^\mu\rangle = \langle\frac{\beta}{8g_s}{\tilde G}^2\rangle = \nonumber \\
%\frac{5}{2}\frac{N(N^2-1)}{128\pi^4}\Lambda^4  \nonumber \\
&&
%\frac{5}{2}\frac{N(N^2-1)}{128\pi^6}
\frac{N^2-1}{64\pi^3}
%\Lambda^4
\frac{M^4C^2}{\left(1+\frac{|h|^2}{M^2}\frac{\beta(\pi)}{4\pi}\right)^2}\exp\left(\frac{4}{\beta_0\pi}\left(1+\frac{|h|^2}{M^2}\frac{\beta(\pi)}{4\pi}\right)\right).
\eea
This potential is characterized by two scales: The Higgs scale given by
\be
h_0^2=\frac{4\pi M^2}{\beta(\pi)},
\ee
and a general scale given by
% Fixed after referee's comments
\be
%V_0=\frac{5}{2}\frac{N(N^2-1)}{128\pi^8}\frac{M^4}{C}.
V_0=\frac{N^2-1}{64\pi^3}M^4C^2.
%\frac{5}{2}\frac{N(N^2-1)}{128\pi^6}M^4C^2.
\ee
Therefore, the form of the potential will be,
\be \label{eqn:pot}
%V({\bar h})=V_0\frac{1}{1+{\bar h}^2}\exp\left(\frac{2}{\beta_0}(1+{\bar h}^2)\right),
V(|{\bar h}|)=V_0\frac{1}{\left(1+|{\bar h}|^2\right)^2}\exp\left(\frac{4}{\beta_0\pi}(1+|{\bar h}|^2)\right),
\ee
where we have introduced the new field ${\bar h}=h/h_0$. This kind of potential has the behavior depicted in Fig.~\ref{fig:pote}.
% Fixed after referee's comments
%It is seen that there is a plateau in 0, granting a possible meta-stable state, and two minima are seen granting the breaking of the electro-weak symmetry.
% Removed after referee's comments
% To have a magnitude order for $V_0$ and $h_0$, we set $M^4/C=1$, $N_c=3$ and kept $\alpha_s=\pi$ at the infrared fixed point as happens for QCD. We get $V_0=4.3227\cdot 10^{-4}\ \text{TeV}$ and $\bar{h}=1.8433\ \text{TeV}$. In such a case, the minima of the potential are attained for $V=\pm 0.67\ \text{TeV}$, in the right ballpark for the electroweak breaking.
%
% Added after referee's comments
We can map the potential in eq.(\ref{eqn:pot}) to the SM Higgs potential that has the parameters experimentally constrained. We get the following approximation for ${\bar h}\ll 1$,
\be
V(|{\bar h}|)=V_0e^{\frac{4}{\beta_0\pi}}-\frac{2V_0}{\beta_0\pi}e^{\frac{4}{\beta_0\pi}}(-2+\beta_0\pi)|{\bar h}|^2
+\frac{V_0}{\beta_0^2\pi^2}(8-8\beta_0\pi+3\beta_0^2\pi^2)e^{\frac{4}{\beta_0\pi}}|{\bar h}|^4+O(|{\bar h}|^6).
\ee
We can now undo the normalization obtaining
\bea
V(|h|)&=&\frac{N^2-1}{64\pi^3}M^4C^2e^{\frac{4}{\beta_0\pi}}
-\frac{N\beta(\pi)}{16\pi^4(N^2-1)}
M^2C^2
e^{\frac{4}{\beta_0\pi}}
(-2+\beta_0\pi)|h|^2 \nonumber \\
&&+\frac{N^2\beta^2(\pi)}{16\pi^5(N^2-1)^3}C^2(8-8\beta_0\pi+3\beta_0^2\pi^2)e^{\frac{4}{\beta_0\pi}}|h|^4+O(|h|^6).
\eea
So, we see that we have two arbitrary constants, $M$ and $C$, that can fix the parameters for the two terms in the series (omitting the constant at the leading order as inessential), granting agreement  with the Higgs potential.
%
% Added after referee's comments
We can compare this effective potential, by accounting for the $h^6$ contribution, to SMEFT studies to check for consistencies \cite{Falkowski:2023hsg,Bartocci:2024fmm}. For our considerations, we omit the initial constant having no physical meaning and use Warsaw basis for the Wilson coefficients \cite{Falkowski:2023hsg}:
\bea
&&\lambda = -\mu^2_H C_{HD}, \nonumber \\
&&C_H = 2\lambda C_{HD},
\eea
where $\mu_H^2$ is the coefficient of quadratic term in the Higgs potential, $\lambda$ is the coefficient of the quartic term in the Higgs potential and $C_H$ is the Wilson coefficient of the order six operator in the SMEFT. We see that, in our case, we have to bound the Wilson coefficient $C_{HD}$ as given in Ref.~\cite{Bartocci:2024fmm}. By a direct comparison we can write
\bea
&&\mu^2_H=
-\frac{M^{4} C^{2} {\mathrm e}^{\frac{32 N}{\left(N^{2}-1\right)^{2}}} N}{2 \left(N^{2}-1\right) \pi^{3} \mathit{h0}}+\frac{\left(N^{2}-1\right) M^{4} C^{2} {\mathrm e}^{\frac{32 N}{\left(N^{2}-1\right)^{2}}}}{32 \pi^{3} \mathit{h0}},
%\beta(\pi)e^{\frac{32 N}{\left(N^2-1\right)^2}}\frac{C^2M^2}{128\pi^4(N^2-1)}\left(N^4-2 N^2-16 N+1\right), 
\nonumber \\
&&\lambda=
\frac{8 M^{4} C^{2} {\mathrm e}^{\frac{32 N}{\left(N^{2}-1\right)^{2}}} N^{2}}{\left(N^{2}-1\right)^{3} \pi^{3} \mathit{h0}^{2}}-\frac{M^{4} C^{2} {\mathrm e}^{\frac{32 N}{\left(N^{2}-1\right)^{2}}} N}{\left(N^{2}-1\right) \pi^{3} \mathit{h0}^{2}}+\frac{3 \left(N^{2}-1\right) M^{4} C^{2} {\mathrm e}^{\frac{32 N}{\left(N^{2}-1\right)^{2}}}}{64 \pi^{3} \mathit{h0}^{2}},
%\beta^2(\pi)e^{\frac{32 N}{\left(N^2-1\right)^2}}\frac{C^2(N^2-1)}{64 \pi ^3} \left(\frac{3}{16\pi^2}-\frac{4N}{\pi^2\left(N^2-1\right)^2}+\frac{32N^2}{\pi^2(N^2-1)^4}\right)
\nonumber \\
&&C_H=
-\frac{256 M^{4} C^{2} {\mathrm e}^{\frac{32 N}{\left(N^{2}-1\right)^{2}}} N^{3}}{3 \left(N^{2}-1\right)^{5} \pi^{3} \mathit{h0}^{3}}+\frac{16 M^{4} C^{2} {\mathrm e}^{\frac{32 N}{\left(N^{2}-1\right)^{2}}} N^{2}}{\left(N^{2}-1\right)^{3} \pi^{3} \mathit{h0}^{3}}-\frac{3 M^{4} C^{2} {\mathrm e}^{\frac{32 N}{\left(N^{2}-1\right)^{2}}} N}{2 \left(N^{2}-1\right) \pi^{3} \mathit{h0}^{3}}+\frac{\left(N^{2}-1\right) M^{4} C^{2} {\mathrm e}^{\frac{32 N}{\left(N^{2}-1\right)^{2}}}}{16 \pi^{3} \mathit{h0}^{3}}.
%-\beta^3(\pi)e^{\frac{32 N}{\left(N^2-1\right)^2}}\frac{C^2\left(N^2-1\right)}{64M^2\pi^3}\left(-\frac{1}{16\pi^3}+\frac{3N}{2 \pi^3\left(N^2-1\right)^2}-\frac{16N^2}{\pi^3\left(N^2-1\right)^4}+\frac{256N^3}{3\pi^3\left(N^2-1\right)^6}\right). 
\nonumber
\eea
From these, we derive
\be
C_{HD}=\frac{C_H}{2\lambda}\stackrel{N\rightarrow 3}{=}\frac{19}{36h_0^2},
% \frac{\beta(\pi)}{6 \pi  M^2 \left(N^2-1\right)^2 }\frac{A_1}{A_2}, \\
% &&A_1=3 N^{12}-18 N^{10}-72 N^9+45 N^8+288 N^7+708 N^6-432 N^5-1491 N^4 \nonumber \\
% &&-3808 N^3+750 N^2-72 N+3, \nonumber \\
% &&A_2=3 N^8-12 N^6-64 N^5+18 N^4+128 N^3+500 N^2-64 N+3. \nonumber
\ee
and
\be
v=\frac{\mu_H}{\sqrt{\lambda}}\stackrel{N\rightarrow 3}{=}\frac{2}{3}h_0=\frac{4}{3}\frac{\sqrt{\pi}M}{\sqrt{\beta(\pi)}}.
\ee
Consistency reasons yield $\beta(\pi)\gg 16\pi/9$ for $M=O(\text{TeV})$ that agrees fairly well with Fig.~\ref{fig:beta} for eq.(\ref{eq:CS}). This also provides a physical interpretation for the mass scale $M$ as the new physics energy scale. Finally, our lower bound for $C_{HD}$ is about $0.235/M^2$ from the consistency limit for $\beta(\pi)$, well in the range given in Ref.~\cite{Bartocci:2024fmm}.

% Changed after referee's comments
%\section{Glueball Spectrum of the Theory}
\section{Mass Spectrum of the Theory}
\label{sec5}

%\ag{Add \& estimate the spectrum for dark glueball.}

\begin{figure}[H]
\centering
\includegraphics[height=7cm,width=8cm]{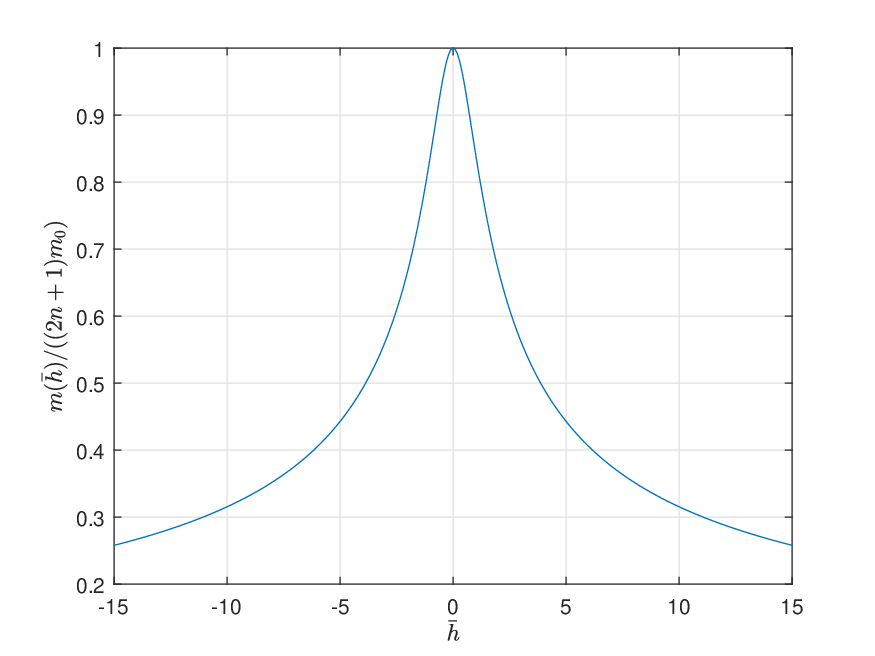}
\caption{\it Behavior of the 
% Changed after referee's comments
%glueball
mass
spectrum as a function of ${\bar h}$. 
}
\label{fig:spec}
\end{figure}

% Changed after referee's comments
The mass spectrum 
%of the glueball 
%
can be evaluated in a standard way as seen in \cite{Frasca:2017slg}. If we are permitted to neglect the mass renormalization shift, that we can assume to be small, the spectrum can be approximated by, for SU(N),
\be
m_n=(2n+1)\frac{\pi}{2K(i)}\left[2\pi N\alpha_s(h)\right]^\frac{1}{4}\mu,
\ee
where $K(i)$ is the complete elliptic integral of the first kind, 
% Added after referee's comment
$\alpha_s(h)$ evaluated at the fixed point
and $\mu$ an integration constant of the theory \cite{Frasca:2015yva}. Therefore, we can take this formula to evaluate the spectrum at the breaking of the symmetry $|h|=v$. Therefore, by using eq.(\ref{eq:alphah}), we get
\be \label{glueball_spectrum}
m_n({\bar h})=(2n+1)m_0\left(1+{\bar h}^2\right)^{-\frac{1}{4}}.
\ee
After having normalized to $(2n+1)m_0$, we get the plot in Fig.~\ref{fig:spec}.
The spectrum is dampened out at increasing values of the Higgs v.e.v.

%\ag{Can we give a plot for the dark glueball spectrum ?}

\medskip

\section{Conclusions and Discussions}
\label{conc}

%\subsection{Summary of our Results}
In this paper, we have investigated a generic Higgs field coupled to SU(N) gauge fields and then dynamically receiving a vacuum expectation value (vev) due to the condensation of the gauge fields below the confinement scale of the strongly coupled SU(N) dark sector. Following a novel technique developed by \textit{Bender et al.}, we were able to compute  this vev analytically as shown in Eqn.~(\ref{eqn:condensate-2}). We also computed the exact RGE of the scalar-gauge dynamics and found the $\beta$-function of theory (Eqn.~(\ref{eqn:running})). Such gauge field condensation and confining dynamics lead to an effective scalar potential of the Higgs field that mimics the standard Higgs potential with the negative mass squared term. We have found the full 
% Changed after referee's comments
%glueball 
mass
spectrum of the theory in Eqn.~(\ref{glueball_spectrum}). We have shown that the potential arising from the coupling with the gauge fields exhibits proper minima that can give rise to electroweak scale breaking (see Fig.~\ref{fig:pote}) while the 
% Changed after referee's comments
% glueball 
mass
spectrum can be dampened out depending on the position of the vev of the Higgs field (see Fig.~\ref{fig:spec}).

%\ag{Add one line of discussion following Fig. 1 and Fig. 2}

%\subsection{Dark Sector Scale Generation}

Such dynamically generated scales due to condensation of gauge fields, with novel mass spectrum of the theory (excited states) and the RGE
% Removed after referee's comment
%, our results are applicable for any BSM model-building. For example, one %may
%
could help to
envisage a SM $\times$ SU(2)$_{D}$ model,  where the Higgs vev which breaks the SU(2)$_{D}$ may have its origin due to strongly coupled gauge sector and the symmetry breaking scale can be very high in general and involve very interesting dark matter physics \cite{Holthausen:2009uc,Hambye:2013dgv,Antipin:2014qva}.   

Finally, we envisage that no scales are special in nature including the EW scale or the Planck scale and entertain the possibility of some dynamical explanation for the generation of scales in nature. So starting from classically scale-invariant theories and scale generation  via non-perturbative phase transition (just like QCD) provides an interesting avenue to understand why different interactions (for example, gravity and EW) behave differently. Such a study involving detailed generation of Planck and EW scales will require deeper investigation and is beyond the scope of the present paper, so we leave it for future studies.

\medskip

\section{Acknowledgement}
\label{Asck}

This work is supported in part by the United States, Department of Energy Grant No. DE-SC0012447 (N.O.).

\newpage

\section*{Appendix A: Dyson-Schwinger Equations \& Bender-Milton-Savage Technique}
\label{AppendixA}

% here below the famous Bender-Milton-Savage technique is presented as a short review \cite{Bender:1999ek}. This allows us to obtain full hierarchy of the Dyson-Schwinger equations in a partial-differential-equation form.

In this Appendix, we briefly present the Bender-Milton-Savage technique \cite{Bender:1999ek}. 
%We consider the partition function for a scalar field
We start from the following partition function for a scalar field
\begin{equation}
    Z[j]=\int[D\phi]e^{iS(\phi)+i\int d^4xj(x)\phi(x)}.
\end{equation}
The equation for the 1P-function is given by
\be
\left\langle\frac{\delta S}{\delta\phi(x)}\right\rangle=j(x)
\ee
where
\be
\left\langle\ldots\right\rangle=\frac{\int[D\phi]\ldots e^{iS(\phi)+i\int d^4xj(x)\phi(x)}}{\int[D\phi]e^{iS(\phi)+i\int d^4xj(x)\phi(x)}}.
\ee
% Then we set $j=0$ and derive the above equation for the 1P-function and dependent on $j$ to obtain the equation for the 2P-function. The definition of the nP-function is given by
One has
\be
\langle\phi(x_1)\phi(x_2)\ldots\phi(x_n)\rangle=\frac{\delta^n\ln(Z[j])}{\delta j(x_1)\delta j(x_2)\ldots\delta j(x_n)},
\ee
and
\be
\frac{\delta G_k(\ldots)}{\delta j(x)}=G_{k+1}(\ldots,x).
\ee

%This means that, for a $\phi^4$ theory, one has
We apply all this to a $\phi^4$ theory and get
\be
S=\int d^4x\left[\frac{1}{2}(\partial\phi)^2-\frac{\lambda}{4}\phi^4\right],
\ee
so that,
\be
\label{eq:G_1}
\partial^2\langle\phi\rangle+\lambda\langle\phi^3(x)\rangle = j(x).
\ee
%The following equation holds to be true
For the 1P-function is
\be
Z[j]\partial^2G_1^{(j)}(x)+\lambda\langle\phi^3(x)\rangle = j(x),
\ee
%By the definition of the 1P-function one gets
and by definition
\be
Z[j]G_1^{(j)}(x)=\langle\phi(x)\rangle.
\ee
%Now we execute the derivative with respect to $j(x)$ to obtain
After derivation with respect to $j(x)$, we get
\be
Z[j][G_1^{(j)}(x)]^2+Z[j]G_2^{(j)}(x,x)=\langle\phi^2(x)\rangle,
\ee
and after another derivation step it one has:
\be
Z[j][G_1^{(j)}(x)]^3+3Z[j]G_1^{(j)}(x)G_2(x,x)+Z[j]G_3^{(j)}(x,x,x)=\langle\phi^3(x)\rangle.
\ee
%Inserting it into Eqn.(\ref{eq:G_1}) yields
We insert this into Eqn.(\ref{eq:G_1}) to obtain
\be
\label{eq:G1_j}
\partial^2G_1^{(j)}(x)+\lambda[G_1^{(j)}(x)]^3+3\lambda G_2^{(j)}(x,x)G_1^{(j)}(x)+G_3^{(j)}(x,x,x)=Z^{-1}[j]j(x)
\ee
% We realize that, via the effects of renormalization, a mass term has appeared. Therefore, setting $j=0$, one gets the first Dyson-Schwinger equation into differential form
Setting $j=0$, we get the first Dyson-Schwinger equation
\be
\partial^2G_1(x)+\lambda[G_1(x)]^3+3\lambda G_2(x,x)G_1(x)+G_3(x,x,x)=0,
\ee
where we realize that quantum corrections induced a mass term.

%By deriving Eqn.(\ref{eq:G1_j}) again with respect to $j(y)$, we see
Our next step is to derive Eqn.(\ref{eq:G1_j}) again with respect to $j(y)$. This gives the result
\be
\begin{split}
&\partial^2G_2^{(j)}(x,y)+3\lambda[G_1^{(j)}(x)]^2G_2^{(j)}(x,y)+
\nonumber \\
&3\lambda G_3^{(j)}(x,x,y)G_1^{(j)}(x)
+3\lambda G_2^{(j)}(x,x)G_2^{(j)}(x,y)
+G_4^{(j)}(x,x,x,y)=\nonumber \\
&Z^{-1}[j]\delta^4(x-y)+j(x)\frac{\delta}{\delta j(y)}(Z^{-1}[j]).
    \end{split}
\ee
%Inserting $j=0$, the equation for the 2P-function becomes
By setting $j=0$, one has for the 2P-function
\be
\partial^2G_2(x,y)+3\lambda[G_1(x)]^2G_2(x,y)+
3\lambda G_3(x,x,y)G_1(x)
+3\lambda G_2(x,x)G_2(x,y)
+G_4(x,x,x,y)=
\delta^4(x-y).
\ee
%This procedure can be iterated to any arbritrary order providing all the hierarchy of Dyson-Schwinger equations in PDE form.
In principle, one can iterate such a procedure to whatever desired order yielding the full hierarchy of Dyson-Schwinger equations into PDE form.

% Added after referee's comments
\section*{Appendix B: Gluon condensate}
\label{AppendixB}

We have to evaluate
\be
\langle G_{\mu\nu}G^{\mu\nu}\rangle = \langle\left(\partial_\mu A_\nu^a-\partial_\nu A_\mu^a+gf^{abc}A_\mu^bA_\nu^c\right)
\left(\partial^\mu A^{a\nu}-\partial^\nu A^{a\mu}+gf^{ade}A^{d\mu}A^{e\nu}\right)
\rangle,
\ee
using our exact solution for the equations of motion and the Green function presented in Appendix C. We have seen that these kind of solutions are obtained by applying a mapping theorem on a quartic scalar field writing $A_\mu^a(x)=\eta_{\mu}^a\phi(x)$, where $\eta_\mu^a$ are constants and $\phi(x)$ the solution for the scalar field equation of motion.
%the computation is accomplished by using eq.(\ref{eq:phisol1}) and the propagator (\ref{eq:G2sol}). 
After some algebra, 
%using eq.(\ref{eq:map}), 
we get
\bea
&&\langle G_{\mu\nu}G^{\mu\nu}\rangle = \langle\left(\partial_\mu A_\nu^a-\partial_\nu A_\mu^a)\right)\left(\partial^\mu A^{a\nu}-\partial^\nu A^{a\mu}\right) \rangle+ \nonumber \\
&&2gf^{ade}\langle\left(\partial_\mu A_\nu^a-\partial_\nu A_\mu^a\right)A^{d\mu}A^{e\nu}\rangle 
+g^2f^{abc}f^{ade}\langle A_\mu^bA_\nu^cA^{d\mu}A^{e\nu}\rangle \nonumber \\
&&= 2\eta_\nu^a\eta^{a\nu}\langle(\partial\phi)^2\rangle-2\eta_\mu^a\eta^{a}_\nu\langle\partial^\mu\phi\partial^\nu\phi\rangle
+g^2f^{abc}f^{ade}\eta_\mu^b\eta_\nu^{c}\eta^{d\mu}\eta^{e\nu}\langle\phi^4\rangle,
\eea
where we have usedf the anti-symmetry of $f^{abc}$ that removes the mixed term linear in $g$. Now, we use the properties of the $\eta$-symbols
\be
\eta_\mu^a\eta^{\mu b}=D\delta^{ab}, \quad \eta^a_\mu\eta^a_\nu=(N^2-1)g_{\mu\nu},
\ee
to obtain
\be
\langle G_{\mu\nu}G^{\mu\nu}\rangle = 2(D-2)(N^2-1)\langle(\partial\phi)^2\rangle+4\pi DN(N^2-1)\alpha_s\langle\phi^4\rangle,
\ee
where $D$ is the number of space-time dimensions that in our case is 4 and the identities $f^{abc}f^{abd}=N\delta^{cd}$ and $f^{abc}f^{abc}=N(N^2-1)$ have been used. In order to evaluate the gluon condensate, we observe that the scalar field that maps the Yang-Mills theory should have the partition function
\be
Z[j] = \int [d\phi]e^{4i(N^2-1)\int d^4x\left[\frac{1}{2}(\partial\phi)^2-\frac{Ng^2}{4}\phi^4+j\phi\right]}
\ee
and we are able to get an explicit expression for it using just the corresponding 1P- and 2P-correlation functions. We can write
\be
Z[j]={\cal N}e^{i\int d^4x G_1(x)j(x)+i\int d^4xd^4x'j(x)G_2(x-x')j(x')+\ldots},
\ee
and
\bea
&&\langle\partial_\mu\phi(x_1)\partial^\mu\phi(x_2)\rangle=\left.\partial_\mu\frac{\delta}{\delta j(x_1)}\partial^\mu\frac{\delta \ln Z}{\delta j(x_2)}\right|_{j=0}, \nonumber \\
&&\langle\phi_(x_1)\phi(x_2)\phi(x_3)\phi(x_4)\rangle =\left.\frac{\delta^4\ln Z}{\delta j(x_1)\delta j(x_2)\delta j(x_3) \delta j(x_4)}\right|_{j=0},
\eea
from which we derive
\bea
&&\langle(\partial\phi)^2\rangle=i\partial^2G_2(x,x), \nonumber \\
&&\langle\phi^4\rangle=iG_4(x,x,x,x),
\eea
but $G_4(x,x,x,x)$ is 0 for our Gaussian solution being a higher-order correlation function evaluated in one single point $x$. This yields
\be
\langle(\partial\phi)^2\rangle=\int\frac{d^4p_E}{(2\pi)^4}p_E^2\sum_{n=0}^\infty\frac{B_n}{p^2_E+m_n^2},
\ee
where the subscript $E$ just means that we moved to Euclidean momentum. The integral needs to be regularized and we use a cut-off $\Lambda$ for this purpose. Taking $\Lambda$ large enough, we are left with
\be
\langle(\partial\phi)^2\rangle=\frac{\Lambda^4}{32\pi^2},
\ee
where use has been mad of the identity $\sum_{n=0}^\infty B_n=1$. This yields for the gluon condensate
\be
\langle\frac{\alpha_s}{\pi} G_{\mu\nu}G^{\mu\nu}\rangle = \alpha_s\frac{N^2-1}{8\pi^3}\Lambda^4.
\ee
%

% Added after referee's comments
\section*{Appendix C: Exact propagator for Yang-Mills theory}
\label{AppendixC}

% Completely rewritten after referee's comment
% Both the propagators for the gluon and the ghost fields can be obtained by selecting some given 1P-correlation functions. This will break the translation invariance but the reason for this is that the vacuum is properly described by a Fubini instanton \cite{Fubini:1976jm,Lipatov:1976ny}. This grants the existence of a mass gap for the theory that, anyway, both the field and 1P-correlation functions are not observables, differently from the Abelian case. This means that there is no experimental way to detect such a violation and a zero mode appears even considering that the 2P-correlation function recover translation invariance and it is this function that enters into the computations of cross-sections and decay rates as per LSZ theorem.

We derive the propagator for Yang-Mills theory by assuming that the vacuum of the theory is well-described by a Fubini instanton \cite{Fubini:1976jm,Lipatov:1976ny} that breaks translation invariance. Such a breaking would be a problem if some experiment could be devised to measure the 1P-correlation function but it is well-known that neither the potentials nor the fields are observable in non-Abelian gauge theories. On the other hand, we will show how the propagator recovers translation invariance and so, when decay rates or cross sections are computed using LSZ decomposition, there is no way to detect breaking of such a symmetry. This kind of 1P-correlation function grants the existence of a mass gap for the theory but this choice is not unique. In our case, it appears to give results in agreement with lattice data.

We want to solve the eq.(\ref{eq:G1j}) and (\ref{eq:G2j}) using a Fubini instanton.
% Changed after referee's comments
%We are granted of this kind of solution by taking 
We take an ansatz that makes all (dark) colour fields proportional to a
single scalar one with proportionality constants carrying the colour
and Minkowski indices \cite{Frasca:2007uz,Frasca:2009yp,Frasca:2015yva}
\be
\label{eq:map}
G_{1\mu}^a(x)=\eta_\mu^a\phi(x),
\ee
where $\eta_\mu^a$ are such a set of constants
% Added after referee's comment
(vacuum impedances)
%
% with mixed indexes 
and $\phi(x)$ a scalar field for a quartic potential. Similarly, we can take
\be
G_{2\mu\nu}^{ab}(x,y)=\delta_{ab}\left(g_{\mu\nu}-\frac{\partial_\mu\partial_\nu}{\partial^2}\right)G_2(x,y),
\ee
where we wrote the gauge projector using the metric $g_{\mu\nu}$ and the derivatives $\partial_\mu$ and now $G_2(x,y)$ is the propagator of the $\phi$ field. By a direct substitution into eq.(\ref{eq:G1j}) and (\ref{eq:G2j}), we get \cite{Frasca:2015yva}
\bea\label{eq:eom}
&&\partial^2\phi+\delta m^2\phi+Ng^2\phi^3=0, \\ \nonumber
&&\partial^2 G_2(x,y)+\delta m^2G_2(x,y)+3Ng^2\phi^2(x)G_2(x,y)=\delta^4(x-y),
\eea
where 
% Changed after referee's comment
\be\label{eq:dm2}
\delta m^2=2Ng^2 G_2(x,x),
\ee
is mass shift arising from quantum effects and in need for regularization being divergent. After renormalization, we know that such a shift has a very small effect on the spectrum of the theory \cite{Frasca:2017slg} and we do not consider it here and in the rest of the paper. Anyway, we observe that the field gains mass simply due to quantum fluctuations but a different choice of the $G_1$ function being a constant fails to yield a mass gap \cite{Chatterjee:2024dgw}. The choice in eq.(\ref{eq:map}) grants that the ghost field decouples and behaves as a free field and this will yield the propagator
\be
K_2(p)=-\frac{1}{p^2+i\epsilon}.
\ee
% Added after referee's comment
This is consistent with the decoupling solution observed in lattice computation \cite{Bogolubsky:2007ud,Cucchieri:2007md,Oliveira:2007px} where the ghost propagator is seen to be the same of a free massless particle with properly changed sign.

The solutions are obtained by noting that the equation for $\phi$ can be solved by taking\footnote{This can be checked by direct substitution and using the properties of the Jacobi elliptical functions.}
\be
\label{eq:phisol1}
\phi(x)=\mu\left(\frac{2}{Ng^2}\right)\operatorname{sn}(p\cdot x+\theta,-1),
\ee
sn being a Jacobi elliptical function and provided that
\be
\label{eq:disp}
p^2=\sqrt{\frac{\lambda}{2}}\mu^2,
\ee
and $\mu$ and $\theta$ are arbitrary integration constants. Then, the equation for $G_2$ can be solved, maintaining the same approximation to neglect $\delta m^2$ 
% Added after referee's comment
in eq.(\ref{eq:eom}) given by eq.(\ref{eq:dm2}), 
by taking for the homogeneous equation 
\be
\partial^2 G_{2h}(x,y)+3Ng^2\phi^2(x)G_{2h}(x,y)=0,
\ee
and the solution can be written as (after a redefinition of the phase $\theta$)\footnote{We are permitted to do so and this can also be seen approaching the problem differently by choosing ${\bm p}=0$ in eq.(\ref{eq:phisol1}), this being an arbitrary parameter again, and solving in $t$ instead, the presence of the phase being inessential for the solution of the homogeneous equation.}
\begin{equation}
G_{2h}(x,y)=a\operatorname{cn}(p\cdot (x-y)+\theta,-1)\operatorname{dn}(p\cdot (x-y)+\theta,-1),
\end{equation}
where $\operatorname{cn}$ and $\operatorname{dn}$ are Jacobian elliptic
functions. One has
\begin{equation}
\frac{d}{dz}\operatorname{sn}(z,i)=\operatorname{cn}(z,i)\operatorname{dn}(z,i)
\end{equation}
and
\begin{equation}
\operatorname{sn}(z,i)=\frac{2\pi}{K(i)}\sum_{n=0}^\infty(-1)^n
  \frac{e^{-\left(n+\frac12\right)\pi}}{1+e^{-(2n+1)\pi}}
  \sin\left((2n+1)\frac{\pi z}{2K(i)}\right).
\end{equation}
Such a solution can be cast in a Fourier series,
\begin{equation}
G_{2h}(x,y)=\frac{\pi^2}{K^2(i)}\sum_{n=0}^\infty(-1)^n(2n+1)
  \frac{e^{-\left(n+\frac12\right)\pi}}{1+e^{-(2n+1)\pi}}
  \cos\left((2n+1)\frac{\pi}{2K(i)}(p\cdot (x-y)+\theta)\right).
\end{equation}
The choice $\theta=(4m+1)K(i)$ with $m\in\mathbb{Z}$ for the phase grants that
the function $\phi_0(x)$ is zero on the light cone. 
% In quantum field theory
% this solution would give immediately the Feynman propagator. In classical
% field theory, we have to use the Laplace--Fourier transform to get the initial
% conditions right. In order to work out this aim, we calculate
Using the Laplace--Fourier transform
\begin{equation}
{\tilde G}_{2h}(\omega,{\bm x}-{\bm y})=\int_0^\infty dt e^{-i(\omega+i\epsilon)t}G_{2h}(x,y),
\end{equation}
where $\epsilon>0$ is a small quantity which is sent to zero in the end in order
to obtain the correct (Feynman) integration path. After integration, we get
\begin{equation}
G_{2h}(\omega,{\bm x}-{\bm y})=\frac12\sum_{n=0}^\infty A_n
  \left[\frac{e^{i(2n+1)\frac{\pi}{2K(i)}{\bm p}\cdot({\bm x}-{\bm y})}}{(\omega
  -(2n+1)\frac{\pi}{2K(i)}p_0)+i\epsilon}-\frac{e^{-i(2n+1)\frac{\pi}{2K(i)}{\bm p}
  ({\bm x}-{\bm y})}}{(\omega+(2n+1)\frac{\pi}{2K(i)}p_0)+i\epsilon}\right],
\end{equation}
where
\begin{equation}
A_n=\frac{\pi^2}{K^2(i)}(2n+1)
  \frac{e^{-\left(n+\frac{1}{2}\right)\pi}}{1+e^{-(2n+1)\pi}}.
\end{equation}
We can perform a Fourier transform in space and use the dispersion relation (\ref{eq:disp}) to obtain
\begin{eqnarray}
G_{2h}(\omega,{\bm k})&=&\frac12\sum_{n=0}^\infty A_n
  \Bigg[\frac{\delta^3({\bm k}+(2n+1)\frac{\pi}{2K(i)}{\bm p})}{(\omega
  -(2n+1)\frac{\pi}{2K(i)}\sqrt{{\bm p}^2+m^2})+i\epsilon}\nonumber\\&&\strut
  -\frac{\delta^3({\bm k}-(2n+1)\frac{\pi}{2K(i)}{\bm p})}{(\omega
  +(2n+1)\frac{\pi}{2K(i)}\sqrt{{\bm p}^2+m^2}))+i\epsilon}\Bigg].
\end{eqnarray}
Finally, we integrate out the arbitrary three-momentum $p$ in order to get to
the Feynman propagator (which is explicitly given only in full momentum
space), using the covariant integration
$\int d^3p/2E_p$ with $E_p=\sqrt{{\bm p}^2+m^2}$, to obtain
\begin{eqnarray}
G_{2h}(\omega,{\bm k})&=&\frac{1}{4\sqrt{{\bm k}^2/((2n+1)\frac{\pi}{2K(i)})^2+m^2}}
  \sum_{n=0}^\infty A_n\times\strut\nonumber\\&&\strut
  \Bigg[\frac{1}{(\omega-(2n+1)\frac{\pi}{2K(i)}\sqrt{{\bm k}^2/((2n+1)
    \frac{\pi}{2K(i)})^2+m^2})+i\epsilon}\nonumber\\&&\strut
  -\frac{1}{(\omega+(2n+1)\frac{\pi}{2K(i)}\sqrt{{\bm k}^2/((2n+1)
    \frac{\pi}{2K(i)})^2+m^2})+i\epsilon}\Bigg].
\end{eqnarray}
Collecting the terms, we obtain back the formula presented in the main text. We see that the translation invariance is fully recovered for $G_2$.

\medskip

\end{document}